\documentclass[12pt]{article}
\usepackage{amssymb,graphicx,color,cite,epsfig,amsfonts,amsmath}
\begin{document}

\baselineskip=.22in
\renewcommand{\baselinestretch}{1.2}
\renewcommand{\theequation}{\thesection.\arabic{equation}}
~\vspace{12mm}
%\begin{flushright}
%{\tt arXiv:YYMM.NNNN}
%\end{flushright}

\begin{center}
{{{\Large \bf Tachyon Vacuum Solution in Open String Field Theory\\[2mm]
with Constant $B$ Field}
}\\[12mm]

{Akira Ishida${}^{(1)}$,~~Chanju Kim${}^{(2)}$,~~Yoonbai
Kim${}^{(1)}$,
~~O-Kab Kwon${}^{(1)}$,~~D. D. Tolla${}^{(3)}$}\\[6mm]
{\it ${}^{(1)}$Department of Physics, BK21 Physics Research
Division,
and Institute of Basic Science\\
Sungkyunkwan University, Suwon 440-746, Korea}\\
{\tt ishida, yoonbai, okab@skku.edu}\\[5mm]
{\it ${}^{(2)}$Department of Physics, Ewha Womans University,
Seoul 120-750, Korea}\\
{\tt cjkim@ewha.ac.kr}\\[5mm]
{\it ${}^{(3)}$Center for Quantum Space Time (CQUeST), Sogang University\\
Shinsu-dong 1, Mapo-gu, Seoul, Korea}\\
{\tt tolla@sogang.ac.kr} }
\end{center}
\vspace{10mm}

\begin{abstract}
We show that Schnabl's tachyon vacuum solution is an exact solution of
the equation of motion of Witten's open bosonic string field theory
in the background of constant antisymmetric two-form field.
The action computed at the vacuum solution is given by the Dirac-Born-Infeld
factor multiplied to that without the antisymmetric tensor field.
\end{abstract}

%{\it{Keywords}} :

\newpage

\setcounter{equation}{0}
\section{Introduction}\label{section1}
Recently there has been much interest in Witten's open string field theory
(OSFT) for bosonic strings~\cite{Witten:1985cc}.
It is a natural framework to discuss the nonperturbative physics of D-branes
as demonstrated in \cite{Sen:1999nx}.
Several conjectures were made on tachyon condensation in \cite{Sen:1999mh},
and the open-string tachyon was identified with the unstable
mode of the D-brane.
The first conjecture states that the tension of the D-brane
is given by the height of the tachyon potential from the tachyon vacuum.
In OSFT, Schnabl proved it by constructing an analytic tachyon vacuum
solution~\cite{Schnabl:2005gv}. Since then, Schnabl's
method~\cite{Okawa:2006vm,Fuchs:2006hw} has been
further developed to prove Sen's the third conjecture~\cite{Ellwood:2006ba}
as well as to study related problems analytically in OSFT, including rolling
tachyon~\cite{Rastelli:2006ap,
Schnabl:2007az,Kiermaier:2007ba,Erler:2007rh,Okawa:2007ri,Ellwood:2007xr,
Fuchs:2007yy,Kiermaier:2007vu,Erler:2007xt,Lee:2007ns,Kwon:2008ap,
Bagchi:2008et,Hellerman:2008wp}.

In this paper we would like to consider Witten's OSFT in the background of
constant antisymmetric two-form field $B$ and find the tachyon vacuum solution.
This background has attracted attention since, in the presence of the $B$
field, the end points of the open strings become noncommutative and
low energy dynamics of the D-branes is described either by commutative or
noncommutative gauge theories~\cite{Seiberg:1999vs}.
Witten's OSFT in the constant $B$ field has been studied
in~\cite{Sugino:1999qd,Kawano:1999fw,Kawano:2000uq}, and it is verified
that the action has the same form as the corresponding action on trivial
background but it has Moyal type noncommutativity in addition to
the ordinary Witten's star product and string coupling
constant~\cite{Sugino:1999qd,Kawano:1999fw}.
We observe that Schnabl's tachyon vacuum solution~\cite{Schnabl:2005gv} still
satisfies the OSFT equation of motion even in this background.
By direct substitution of the solution into the action we identify the
tension of D25-brane in the presence of constant $B$ field, which
is nothing but the D25-brane tension multiplied by the Dirac-Born-Infeld (DBI)
factor $\sqrt{-\det(\eta + 2\pi\alpha' B)}\,$. Since $B$ field appears
only through the gauge invariant combination $B+F$ where $F$ is the
field strength on the D-brane~\cite{Seiberg:1999vs},
this result is consistent with the DBI action obtained by summing
over all disk diagrams with external constant $F$.

\setcounter{equation}{0}
\section{Constant $B$ Field and Tachyon Vacuum}
\label{section2}

Dynamics of bosonic strings including off-shell contributions is
described by Witten's OSFT~\cite{Witten:1985cc}, in which the action
is
\begin{align}
S(\Phi) =& -\frac{1}{{\alpha'}^3g_{{\rm o}}^2}\int \left(
\frac{1}{2}\,\Phi*Q_{{\rm B}} \Phi + \frac{1}{3}
\Phi*\Phi*\Phi\right)
\label{ac1}\\
= & -\frac{1}{{\alpha'}^3g_{{\rm o}}^2} \left( \frac{1}{2}
~_{12}\langle V_2||\Phi\rangle_1 Q_{{\rm B}} |\Phi\rangle_2
+\frac{1}{3} ~_{123}\langle V_3 ||\Phi\rangle_1\Phi\rangle_2
\Phi\rangle_3\right), \label{ac2}
\end{align}
where $g_{{\rm o}}$, $\Phi$, $\ast$, and $Q_{{\rm B}}$ in the first
line \eqref{ac1} are the open string coupling constant, a string
field, Witten's star product, and the BRST charge, respectively. In
the second line, $~_{12...n}\!\langle V_n|$ is the $n$-string overlap
vertex and $|\Phi\rangle_{n}$ is the $n$-th string state.
From (\ref{ac2}) the equation of motion is
\begin{equation}\label{eom1}
~_{12}\langle V_2 | Q_{{\rm B}} |\Phi\rangle_2 +~_{123}\langle
V_3||\Phi\rangle_2|\Phi\rangle_3 =0.
\end{equation}

The analytic tachyon vacuum solution is known to be~\cite{Schnabl:2005gv}
\begin{equation}\label{ss}
|\Psi\rangle=\lim_{N\to\infty}\left(\sum_{n=0}^N\psi_n' -
\psi_N\right),
\end{equation}
where $\psi_n$ is expressed in terms of ghosts and wedge state $|n\rangle$,
($n=0,1,\ldots$),
\begin{equation}
\psi_n = \frac{2}{\pi} c_1|0\rangle*|n\rangle* B_1^L c_1|0\rangle,
\end{equation}
and $\psi_n'\equiv\frac{d\psi_n}{dn}$.
$B_1^L$ is represented in terms of $b$ ghost as
\begin{equation}
B_1^L =\int_{C_L}\frac{d\xi}{2\pi i} (1+\xi^2) b(\xi),
\end{equation}
where the contour $C_L$ runs counterclockwise along the unit circle
with ${\rm Re}(\xi)<0$. In (\ref{ss}), the coefficient of the so-called
phantom piece $\psi_N$ has to be $-1$ for the solution to satisfy
the equation of motion when contracted with
itself~\cite{Okawa:2006vm,Fuchs:2006hw},
\begin{equation}\label{eom2}
~_{12}\langle V_2 ||\Psi\rangle_1 Q_{{\rm B}} |\Psi\rangle_2
+~_{123}\langle V_3||\Psi\rangle_1 |\Psi\rangle_2 |\Psi\rangle_3 =0.
\end{equation}
Evaluation of the action for the Schnabl's solution gives
\begin{equation}\label{acsnb}
S(\Psi)=\frac{1}{6{\alpha'}^3 g_{\rm o}^2}~_{123} \langle
V_3||\Psi\rangle_1 |\Psi\rangle_2 |\Psi\rangle_3
=\frac{{\rm Vol}_{26}}{2\pi^2{\alpha'}^3 g_{\rm o}^2}
\equiv {\cal T}_{25} {\rm Vol}_{26},
\end{equation}
where ${\rm Vol}_{26} =\int d^{26}x$ represents the spacetime volume
factor and ${\cal T}_{25}$ is the tension of 25-dimensional space-filling brane.
Apart from the volume factor, this coincides with
the tension of D25-brane,
in consistent with Sen's conjecture \cite{Sen:1999mh}.

In the presence of a constant antisymmetric two-form field, the string
worldsheet action is written as
\begin{align}
\frac{1}{2\pi\alpha'}\int d^2z\,(g_{MN}
-2 \pi\alpha'B_{MN})\partial X^{M}
\bar\partial  X^N+S_{{\rm gh}}, \label{WSA}
\end{align}
where $S_{{\rm gh}}$ is the $b,c$ ghost contribution. We denote the
Greek indices $\mu,\ \nu$ as the directions that the antisymmetric tensor field
is nonzero, $B_{\mu\nu}\neq 0$.
The classical solution of the equation of motion under the boundary
condition $E_{\mu\nu}\partial_{\bar z} X^\nu =
(E^{T})_{\mu\nu}\partial_z X^\nu$ with $E_{\mu\nu} = g_{\mu\nu} +
2\pi \alpha' B_{\mu\nu}$ is given by
\begin{eqnarray}\label{Xzz}
X^\mu (z,\bar z)&=& \tilde x^\mu - i\alpha'\left[(E^{-1})^{\mu\nu}
\ln \bar z + (E^{-1 T})^{\mu\nu}\ln z\right]p_\nu
\nonumber \\
&&+ i\sqrt{\frac{\alpha'}{2}}\, \sum_{n\ne 0} \frac{1}{n}
\left[(E^{-1})^{\mu\nu}\alpha_{n,\nu}\bar z^{-n} + (E^{-1
T})^{\mu\nu} \alpha_{n,\nu} z^{-n}\right],
\end{eqnarray}
where $\tilde x^\mu$'s are noncommutative coordinates satisfying
\begin{equation}
[\tilde x^\mu,\, \tilde x^\nu] = i \theta^{\mu\nu}, \qquad
\theta^{\mu\nu} = -(2\pi\alpha')^2 (E^{-1} B E^{-1 T})^{\mu\nu}.
\end{equation}
The commutation relations among operators are
\begin{equation}\label{ctr1}
[\alpha_{n,\, \nu},\, \alpha_{m,\,\nu}] = n G_{\mu\nu}
\delta_{n+m,\,0}, \quad [x^\mu,\, p_\nu]=i\delta^{\mu}_{\,\nu},
\end{equation}
where $G_{\mu\nu}$ is the inverse matrix of $G^{\mu\nu}=(E^{-1}
gE^{-1 T})^{\mu\nu}$ and $x^\mu = \tilde x^{\mu} +
\frac{1}{2}\theta^{\mu\nu} p_\nu$ is the center of mass coordinate.

The corresponding OSFT action in the constant $B$ field background
is~\cite{Sugino:1999qd,Kawano:1999fw}
\begin{align}
S_{B}=& -\frac{1}{{\alpha'}^3G_{\rm o}^2}\int \left( \frac{1}{2}\,
\Phi\star \tilde Q_{{\rm B}}\Phi + \frac{1}{3}
\Phi\star\Phi\star\Phi\right) \nonumber \\
= & -\frac{1}{{\alpha'}^3G_{\rm o}^2} \left( \frac{1}{2}
~_{12}\langle \hat V_2||\Phi\rangle_1 \tilde Q_{{\rm B}}
|\Phi\rangle_2 +\frac{1}{3} ~_{123}\langle \hat V_3
||\Phi\rangle_1|\Phi\rangle_2 |\Phi\rangle_3\right), \label{bac2}
\end{align}
where $G_{\rm o}$ is the open string coupling in the presence of constant $B$
field~\cite{Seiberg:1999vs},
\begin{equation}\label{Gg}
G_{\rm o}^2= g_{\rm o}^2\sqrt{ \frac{\det(g + 2\pi\alpha'B)}{\det g} }\, .
\end{equation}
In $S_{{\rm B}}$, $\star$ in the first line is Witten's star product
modified by the Moyal-type noncommutativity and $~_{12...n}\!\langle \hat V_n|$
in the second line is the $n$-string overlap vertex corresponding to
the product $\star$ in the $B$ field background.
The BRST charge ${\tilde Q}_{{\rm B}}$ is not affected by the $B$ field
except that the metric in ${\tilde Q}_{{\rm B}}$
changes to the open string metric $G_{\mu\nu}$.
The equation of motion of (\ref{bac2}) is
\begin{equation}\label{eqb}
{}_{12}\langle \hat V_2| \tilde Q_{\rm B} |\Phi\rangle_2
+{}_{123}\langle \hat V_3 | |\Phi\rangle_2 |\Phi\rangle_3 = 0,
\end{equation}
where the mode expansion of ${}_{12}\langle \hat V_2|$, $\tilde
Q_{{\rm B}}$, and ${}_{123}\langle \hat V_3 |$ are expressed in
terms of $G_{\mu\nu}$, $\alpha_{n,\mu}$, $p_{\mu}$, $x^{\mu}$,
$b_{n}$, and $c_n$ defined above.

Now, we find a nontrivial homogeneous solution of this equation which
represents the tachyon vacuum. Let us first introduce new
modes~\cite{Sugino:1999qd}
\begin{equation}\label{modetr}
\hat\alpha_n^\mu = \left(E^{-1 T}\right)^{\mu\nu}\alpha_{n,\nu},\quad
\hat p^\mu = \left(E^{-1 T}\right)^{\mu\nu} p_\nu, \quad
\hat x_\mu = E_{\mu\nu} x^\nu.
\end{equation}
Then the commutation relations become those without $B$ field,
\begin{equation}\label{ctr2}
[\hat\alpha_n^\mu,\,\hat\alpha_m^\nu] = n g^{\mu\nu}\delta_{n+m,0},\quad
[\hat x_\nu,\, \hat p^\mu] =i\delta_\nu^{\,\mu},
\end{equation}
and the bilinear form of $\alpha_{n}^{\mu}$
with metric $G_{\mu\nu}$ is replaced by that of the transformed
modes $\hat\alpha^\mu_n$ with metric $g_{\mu\nu}$, i.e.,
\begin{equation}\label{biltr}
G^{\mu\nu}\alpha_{n,\mu}\alpha_{m,\nu} =
g_{\mu\nu}\hat\alpha_n^\mu\hat\alpha_m^\nu,
\end{equation}
where $\alpha_{0,\mu} = \sqrt{2\alpha'}\,p_\mu$.
Then, ${}_{12}\langle \hat V_2|$
and ${}_{123}\langle \hat V_3 |$ are rewritten in
terms of the new modes,
$g_{\mu\nu}$, $\hat\alpha_n^\mu$, $\hat p^{\mu}$, $\hat x_{\mu}$, $b_{n}$,
and $c_n$, and are related with those in the absence of $B$ field
as~\cite{Sugino:1999qd,Kawano:1999fw},
\begin{align} \label{vertices}
{}_{12}\langle \hat V_2|={}_{12}\langle V_2|,\qquad
{}_{123}\langle \hat V_3 |={}_{123}\langle V_3 |
\exp\left(-\frac{i}{2}\sum_{r<s}\theta^{\mu\nu}p_{\mu}^{(r)}p_{\nu}^{(s)}
\right).
\end{align}
Also, $\tilde Q_{{\rm B}}$ has the same form as $Q_{{\rm B}}$ with the
oscillators replaced by the new ones.

With these forms of overlap vertices and the BRST charge, it is not difficult
to see
that the Schnabl's tachyon vacuum solution \eqref{ss}
is still the solution of the equation of motion \eqref{eqb} even in the
presence of constant $B$ field.
For this, we need
to show that \eqref{eqb} holds
when contracted with any state in the Fock space
as well as with the solution itself\cite{Okawa:2006vm,Fuchs:2006hw}.
First note that the kinetic term reduces to that in the absence of the $B$
field due to \eqref{vertices} and \eqref{ctr2}.
Furthermore, since the tachyon vacuum solution is a homogeneous
solution~\cite{Schnabl:2005gv},
we have $p_{\mu}^{(r)}|\Psi\rangle_{r} =0$ and hence the Moyal type
noncommutativity disappears in the cubic term. Then,
the resulting equation coincides with the equation without
the background $B$ field \eqref{eom2} for homogeneous string fields.
This is true for the contraction with either any state in the Fock space or
the string field itself. This completes the proof.

From now on let us discuss physical properties of the solution.
In the absence of the $B$ field, the calculation of the action \eqref{ac1}
for the Schnabl's analytic vacuum solution gives the exact value
of the tension ${\cal T}_{25}$ of 25-dimensional space-filling
brane~\cite{Schnabl:2005gv} as in (\ref{acsnb}).
Since all the $B$ field dependence in the action (\ref{bac2})
appears only through the open string coupling constant $G_{o}$
and the transformed center of mass coordinate $\hat x^\mu$,
the value of the action (\ref{bac2}) at the tachyon
vacuum solution with constant $B$ field gives
\begin{align}
S_{B}(\Phi=\Psi)=& \frac{1}{6\alpha'^{3}G_o^2}
{}_{123}\langle V_{3}||\Psi\rangle_{1}|\Psi\rangle_{2}|\Psi\rangle_{3}
=\frac{1}{2\pi^2{\alpha'}^3 G_o^2}\int d^{26}{\hat x}
\nonumber \\
=&\frac{{\rm Vol}_{26}}{2\pi^2{\alpha'}^3 G_o^2}
|\det(\eta + 2\pi\alpha' B)|
= {\cal T}_{25}{\rm Vol}_{26}
\sqrt{-\det(\eta + 2\pi\alpha' B)}\,,
\label{vac}
\end{align}
where $\eta_{\mu\nu}$ is the flat spacetime metric,
$|\det(\eta + 2\pi\alpha' B)|$ is the Jacobian factor
for the coordinate transformation given in (\ref{modetr}),  and
we used the relation $G_o^2= \sqrt{-\det (\eta + 2\pi\alpha'B)}\, g_o^2$
from (\ref{Gg}).
Since we are considering the case of
a constant $B$ field,
the result \eqref{vac} is consistent with the DBI action
obtained by summing over all the disk diagrams
with external gauge field legs of constant electromagnetic field strength
$F_{\mu\nu}$ as boundary terms~\cite{Fradkin:1985qd}, considering the fact
that $B$ field appears only through the gauge invariant combination
$B_{\mu\nu}+F_{\mu\nu}$.
The result \eqref{vac} is natural because
the Schnabl's tachyon vacuum solution in the presence of constant $B$ field
background is an exact analytic solution of classical SFT equation
and does not involve any off-shell contribution.

\setcounter{equation}{0}
\section{Conclusion}\label{section3}

In this paper, we showed that Schnabl's tachyon vacuum solution
is an exact solution of the string field equation in the presence of
constant $B$ field. Since the effect of the constant $B$ field appears
only in the open string coupling and the Jacobian factor of spacetime
integration, the value of the SFT action is easily computed and gives
the DBI Lagrangian density multiplied by the D25-brane tension and
spacetime volume, which coincides with the result of disc amplitude
computation in string theory.

The obtained result may open a new direction in OSFT including both constant
$B$ field background and marginal deformations of either time or spatial
dependence. It would be intriguing if the treatment of the $B$ field
becomes applicable to the cases with time and spatial dependence, which
lead to higher-derivative corrections~\cite{Coletti:2003ai}. Since
fundamental strings couple minimally to the antisymmetric tensor field
$B$ in string theories, our work on the OSFT in the presence of the constant
$B$ field may contribute to understanding of fundamental strings in the
context of OSFT.

\section*{Acknowledgements}
This work was supported by Astrophysical Research
Center for the Structure and Evolution of the Cosmos (ARCSEC)) and
grant No. R01-2006-000-10965-0 from the Basic Research Program
through the Korea Science $\&$ Engineering Foundation (A.I.,O.K.),
by the Science Research Center Program of the
Korea Science and Engineering Foundation through the Center for
Quantum Spacetime (CQUeST) with grant
number R11-2005-021 and KRF-2006-352-C00010 (C.K.,D.T.), and
by Faculty Research Fund, Sungkyunkwan University, 2007 (Y.K.).

\end{document}